\documentclass[12pt]{article}
\usepackage{epsfig, amsmath, amssymb}
\usepackage{graphicx,psfrag} 
\newcommand{\be}{\begin{equation}}
\newcommand{\ee}{\end{equation}}
\newcommand{\bea}{\begin{eqnarray}}
\newcommand{\eea}{\end{eqnarray}}
\newcommand{\bA}{\begin{array}}
\newcommand{\eA}{\end{array}}
\newcommand{\bc}{\begin{center}}
\newcommand{\ec}{\end{center}}
\newcommand{\al}{\alpha}

\newcommand{\ra}{\rightarrow}
\newcommand{\del}{\partial}

\newcommand{\ie}{{\it i.e.}}
\newcommand{\eg}{{\it e.g.}}

\begin{document}


\begin{titlepage}

\bc

\hfill 
\\         [22mm]

{\Huge Non-conformal brane plane waves \\ [2mm]
and entanglement entropy}
\vspace{16mm}

{\large K.~Narayan} \\
\vspace{3mm}
{\small \it Chennai Mathematical Institute, \\}
{\small \it SIPCOT IT Park, Siruseri 603103, India.\\}

\ec
\medskip
\vspace{40mm}

\begin{abstract}
Following [arXiv:1202.5935 [hep-th]] and [arXiv:1212.4328 [hep-th]],
we study non-conformal brane plane wave backgrounds dual to strongly 
coupled gauge theories with constant energy flux and holographic
entanglement entropy for strip subsystems in them. We find that for
the strip direction along the direction of the energy flux, the finite
cutoff-independent part of entanglement entropy can be estimated in
terms of a dimensionless combination of the energy density and the
strip dimensions, alongwith an effective scale-dependent number of
degrees of freedom. For the strip orthogonal to the flux direction,
there are indications of phase transitions.  We also briefly discuss
NS5-brane backgrounds corresponding to plane wave states in little
string theories.
\end{abstract}

\end{titlepage}



\section{Introduction and summary}

Entanglement entropy has come to be a useful tool in discussions of
nonrelativistic holography, in light of the simple prescription
of Ryu-Takayanagi \cite{Ryu:2006bv,Ryu:2006ef,HEEreview}. This states 
that the holographic entanglement entropy of a subsystem in the 
boundary $d$-dim conformal field theory is $S_A = {1\over 4G_{d+1}}
Area(\gamma_A)$, where $\gamma_A$ is a $(d-1)$-dim minimal surface
bounding the subsystem and extending into the bulk $d+1$-dim anti de
Sitter spacetime. This prescription has been checked extensively with
success. In the context of an explicit gauge/string realization of 
$AdS_{d+1}/CFT_d$, we identify $G_{d+1}={G_{10}\over V_{9-d}}$ after 
dimensionally reducing on the compact $(9-d)$-dim space: 
correspondingly, the codim-2 minimal surface in question wraps this 
compact space too, and should now be thought of as an 8-dim surface.

It is natural to ask how this generalizes to non-conformal theories 
arising on non-conformal D-branes: \cite{Ryu:2006ef} proposed that 
the entanglement entropy now is\ 
$S_A = {1\over 4G_{10}} \int d^8x e^{-2\Phi} \sqrt{g}$ ,\
$\Phi$ being the dilaton and $g$ the string frame metric. This 
appears sensible and \cite{Ryu:2006ef} studied examples vindicating 
this expression.

Some questions arise in this context. 
Firstly, the natural quantity in an M-theory context is the 11-dim
Einstein metric, which under dimensional reduction (to a Type II
string description) on the 11-th circle is again most naturally
related to the 10-dim Einstein metric.  Secondly, the natural
generalization of the Ryu-Takayanagi prescription in any effective
gravity model again involves the lower dimensional Einstein metric. 
Of relevance in this 
context are certain classes of lightlike $AdS\times S$ deformations, 
$AdS$ plane waves \cite{Narayan:2012hk,Singh:2012un} (see also 
\cite{Narayan:2012wn}), which have simple dual microscopic 
interpretations as conformal field theory excited states with 
constant energy flux $T_{++}$. Some of these exhibit deviations 
from the area law \cite{AreaLaw} of entanglement entropy 
holographically: see \cite{Narayan:2012ks} for a systematic 
treatment of strip-subsystems using the covariant formulation \cite{HRT}.
Upon appropriate dimensional reduction, $AdS$ plane waves become 
gravity backgrounds exhibiting hyperscaling violation: the $AdS_5$ 
case exhibits logarithmic behaviour. Such spacetimes arise in 
effective Einstein-Maxwell-scalar theories \cite{HV}
(see also \cite{Dong:2012se,HVstr}) with 
some of them argued to have signatures of hidden Fermi surfaces 
\cite{OTU,Sa}. In this context again, the Einstein metric appears 
natural for holographic entanglement entropy \eg\ \cite{OTU,Dong:2012se}.

This suggests that the natural generalization of the Ryu-Takayanagi
prescription in M-theory should be 
\be\label{EEMth}
S_A = {1\over 4G_{11}} \int d^9x \sqrt{g}
\ee
where $g$ is the spatial Einstein metric induced on the 9-dim surface 
in the 11-dim bulk spacetime. This trivially agrees with the familiar 
expressions for M2- and M5-brane conformal theories after reducing 
on the spheres. Furthermore, a 9-dim minimal surface in M-theory 
must wrap the 11-th circle to have a sensible interpretation as an 
8-dim surface in the Type II description after dimensional reduction 
on the 11-th circle. This implies  
\be\label{EEMst}
S_A^{11}\ \ra\   {1\over 4G_{10}} \int d^8x \sqrt{g^E_{10}}\ 
 =\  {1\over 4G_{10}} \int d^8x\ e^{-2\Phi} \sqrt{g^{st}_{10}}\ .
\ee
The first expression involving the 10-dim Einstein metric 
is, happily, identical to the second one, proposed in 
\cite{Ryu:2006ef}, when we recall that $g^E_{10}=e^{-\Phi/2}g^{st}_{10}$. 
Alternatively, the 11-dim metric is\
$ds_{11}^2 = e^{-2\Phi/3} ds_{10,st}^2 + e^{4\Phi/3} (dx_{11}+A)^2$:
dimensionally reducing (\ref{EEMth}) on the 11-th circle gives
\be
S_A\ra\ \int {d^8x dx_{11}\over 4G_{11}}\ \sqrt{g_{11,11}}\ e^{-4.2\Phi/3} 
\sqrt{g^{st}_{10}}  
= {1\over 4G_{10}} \int d^8x\ e^{-2\Phi} \sqrt{g^{st}_{10}}\ .
\ee
The last expression pertains to Type II string/supergravity and is in 
fact the one proposed in \cite{Ryu:2006ef}, which we have recovered 
from our expression (\ref{EEMth}).

Proposing that the holographic entanglement entropy (\ref{EEMth})
(\ref{EEMst}) be expressed uniformly in terms of the Einstein metric
in 10- or 11-dim also appears consistent with the expectation that the
corresponding expression for entanglement entropy in a lower
dimensional effective gravity theory obtained after dimensional
reduction on some compact space involves the Einstein metric. 
The study of non-conformal brane ground states and associated 
hyperscaling violating spacetimes in \cite{Dong:2012se} in a sense 
vindicates this.

In what follows, we use the above prescriptions to study entanglement
entropy for strip-subsystems in plane wave excited states in
non-conformal $Dp$-brane theories, following
\cite{Narayan:2012hk,Narayan:2012ks}. These bulk spacetimes arise in
certain zero-temperature double scaling limits of boosted black
$Dp$-branes \cite{Singh:2012un}, and are dual to strongly coupled
gauge theories with constant energy flux $T_{++}$. For the strip along
the energy flux direction, the leading divergent part is the same as
for ground states \cite{Ryu:2006ef,Barbon:2008ut}. The finite
cutoff-independent piece \cite{Ryu:2006bv,Ryu:2006ef,CH}
grows with subsystem size, in accord with the intuition that energy
pumped into the system, and thereby entanglement, increases as the
size increases. It can be written as a dimensionless combination of
the energy density $Q$ and the strip length/width, and further
involves an effective scale-dependent number of degrees of freedom
$N_{eff}(l) = N^2 \left({g_{YM}^2N\over l^{p-3}}\right)^{{p-3\over 5-p}}$\ 
which also appeared for ground states \cite{Barbon:2008ut}. We also
find consistency with the expectation that the finite 
cutoff-independent piece decrease under the renormalization group flow
in the $Dp$-brane phase diagram. For the strip orthogonal to the flux
direction, there are indications of phase transitions, constrained
however by the multiple length scales here, unlike the conformal case.
Finally we point out NS5-brane backgrounds describing plane wave
excited states in the dual little string theories.

\section{Plane wave excited states}

\subsection{Conformal theories}

First we consider $AdS_{d+1}$ plane waves arising on the conformal $D3, 
M2, M5$-branes: the bulk spacetimes corresponding to normalizable 
deformations of $AdS_{d+1}\times S$ are \cite{Narayan:2012hk}
\be\label{AdSpwOld}
ds^2 = {R^2\over r^2} (-2dx^+dx^-+dx_i^2+dr^2) 
+ R^2 {\tilde Q} r^{d-2} (dx^+)^2 + R^2 d\Omega^2\ .
\ee
These are likely $\al'$-exact backgrounds. 
The parameter ${\tilde Q}\gg 0$ gives rise to an energy-momentum 
density $T_{++}$ in the boundary CFT. For uniformizing with the 
non-conformal brane description to follow, it is useful to redefine 
the constant in the $g_{++}$ component as\ 
\be\label{QtildeQconf}
{\tilde Q} = Q {G_{d+1}\over R^{d-1}}\ ;\qquad\quad 
{\tilde Q}={Q\over N^2}\ \ (D3), \quad {\tilde Q}={Q\over N^{3/2}}\ \ (M2),
\quad {\tilde Q}={Q\over N^3}\ \ (M5)\ .
\ee
Recalling the $N$-scaling of the entropy of the various conformal 
branes, we see ${\tilde Q}$ is the energy-momentum density per 
(nonabelian) degree of freedom.\ This gives
\be\label{AdSpw}
ds^2 = {R^2\over r^2} (-2dx^+dx^-+dx_i^2+dr^2) 
+ {G_{d+1} Q\over R^{d-3}} r^{d-2} (dx^+)^2 + R^2 d\Omega^2\ ,
\ee
where $G_{d+1}$ is the $d+1$-dim Newton constant, $Q$ is the 
energy-momentum density component $T_{++}$ in the boundary 
theory. These spacetimes are best thought of in the following way: 
start with $AdS_{d+1}$-Schwarzschild black branes boosted with parameter 
$\lambda$\ (see \eg\ \cite{Maldacena:2008wh})
\bea\label{AdSbblc}
&& ds^2 = {R^2\over r^2} \left[-2dx^+dx^-+ {r_0^nr^d\over 2R^{d+n}}
(\lambda dx^++\lambda^{-1} dx^-)^2 + dx_i^2\right]
+ {dr^2\over r^2 (1-{r_0^nr^d\over R^{d+n}})} + R^2d\Omega^2\ ,\nonumber\\
&& \qquad 
r_0^4\sim G_{10}\varepsilon_4\ (D3)\ ,\qquad\ \ 
r_0^6\sim G_{11} \varepsilon_3\ 
(M2)\ , \qquad\ \ r_0^3\sim G_{11} \varepsilon_6\ (M5)\ ,
\eea
where the terms arising from the finite temperature blackening factor\ 
$(1-{r_0^n\over \rho^n})$\  (with $n=4 (D3), 6 (M2), 3 (M5)$) have been 
recast using the usual $({\rho\over R})^{\#}\ra {R\over r}$ coordinate 
transformation (with $r=0$ now being the $AdS$ boundary), and the 
temperature parameter $r_0$ expressed using the energy density 
$\varepsilon$ above extremality in the respective D3, M2 or M5-brane 
theories. Consider the (zero temperature, infinite boost) double 
scaling limit \cite{Singh:2012un}
\be\label{doubscLim}
\lambda\ra\infty,\ \varepsilon_{p+1}\ra 0,\qquad {\rm with} \qquad
{\lambda^2\varepsilon_{p+1}\over 2}\equiv Q = {\rm fixed}\ .
\ee
Of the finite temperature terms, the metric component $g_{++}$ alone 
survives in this limit, and the spacetimes (\ref{AdSbblc}) reduce to 
the $AdS$ plane wave spacetimes (\ref{AdSpw}), after using
\bc
$G_5\sim G_{10} R_{D3}^5\ \ {\rm or}\ \ G_{4,7}\sim G_{11} R_{M2,M5}^{7,4} ,\ \
{\rm with} 
\ \ R_{D3}^4\sim g_sNl_s^4,\ R_{M2}^6\sim Nl_P^6,\ R_{M5}^3\sim Nl_P^3$~.
\ec
The entanglement entropy for ground states ($Q=0$) in the CFTs arising 
on the various conformal (D3, M2, M5) branes with strip-shaped 
subsystems has the form
\be\label{EEgndconf}
S_A\ \sim\ {R^{d-1}\over G_{d+1}} \Big( {V_{d-2}\over\epsilon^{d-2}} - c_d
{V_{d-2}\over l^{d-2}}\Big)\ ,
\ee
where $c_d>0$ is some constant, $l$ the strip width and $V_{d-2}$ the 
longitudinal size (we are interested in the scaling behaviours alone 
throughout this paper, so will not be careful with numerical 
coefficients). The first term represents the area law while the 
second term is a finite cutoff-independent part encoding a 
size-dependent measure of the entanglement. With $Q\neq 0$, we have 
an energy flux in a certain direction: these are nonstatic 
spacetimes, and we use the covariant formulation of holographic 
entanglement entropy \cite{HRT}. Consider the strip to be along 
the flux direction, \ie\ with width along some $x_i$ direction
\cite{Narayan:2012ks}. Then the leading divergent term is the same as
for ground states.  The width scales as\ $l\sim r_*$ and the finite
cutoff-independent piece in these excited states is
\bea
&& \qquad\quad \pm 
\sqrt{Q} V_{d-2} l^{2-{d\over 2}}\ \sqrt{R^{d-1}\over G_{d+1}}\qquad\qquad 
[+:\ d<4,\ \ \ -:\ d>4]\ ;\nonumber\\
&& \sqrt{Q}V_2 N\ \log (lQ^{1/4})\ \ (D3)\ ;\quad 
\sqrt{Q} L \sqrt{l}\ \sqrt{N^{3/2}}\ \ (M2); \quad
-\sqrt{Q} {V_4\over l}\ \sqrt{N^3}\ \ (M5)\ .
\eea
For the M5-theory, we have\ 
${V_4N^3\over l} (-\sqrt{Q\over N^3} + {1\over l^3}) > 0$\  \ie\ the 
finite part of EE for the excited state is larger than that for the 
ground state if\ $l\ll \sqrt{N} Q^{-1/6}$. This may seem a bit surprising: 
perhaps the best interpretation is that these expressions are 
obtained in the large $N$ approximation (where $\sqrt{N} Q^{-1/6}\ra 
\infty$), with corrections at finite $N$.\\
With the strip orthogonal to the flux direction, a phase 
transition was noted \cite{Narayan:2012ks}: for large width $l$, 
there is no connected surface corresponding to a spacelike subsystem.

\subsection{Non-conformal Dp-brane theories}

The string metric and dilaton for $Dp$-brane plane waves (with 
normalizable $g_{++}$-deformations) are 
\bea\label{Dpnullnorm}
&& ds^2_{st} = {r^{(7-p)/2}\over R_p^{(7-p)/2}} dx_{\parallel}^2 + 
{G_{10} Q_p\over R_p^{(7-p)/2}} {(dx^+)^2\over r^{(7-p)/2}} 
+ R_p^{(7-p)/2} {dr^2\over r^{(7-p)/2}} 
+ R_p^{(7-p)/2} r^{(p-3)/2} d\Omega_{8-p}^2\ ,\nonumber\\
&& e^\Phi = g_s \Big({R_p^{7-p}\over r^{7-p}}\Big)^{{3-p\over 4}} ;
\quad g_{YM}^2\sim g_s {\al'}^{(p-3)/2}\ ,\qquad 
R_p^{7-p}\sim g_{YM}^2N {\al'}^{5-p} \sim g_sN {\al'}^{(7-p)/2}\ .\quad
\eea
As for the conformal branes, the $g_{++}$ term here has been obtained 
starting with the non-conformal finite temperature solutions 
\cite{Itzhaki:1998dd}: using 
the coordinate $r$ where $U={r\over\al'}$, the temperature parameter 
being\ $r_0^{7-p}=(U_0\al')^{7-p}\sim G_{10} \varepsilon_{p+1}$ , we have 
the metric component\ 
$g_{tt}=-{r^{(7-p)/2}\over R_p^{(7-p)/2}} (1-{r_0^{7-p}\over r^{7-p}})$. 
Rewriting in terms of lightcone variables $t={x^++x^-\over\sqrt{2}} ,\
x_p={x^+-x^-\over\sqrt{2}}$~, we obtain (\ref{Dpnullnorm}) 
by a zero temperature, infinite boost, double scaling limit\ 
$\lambda\ra\infty, r_0\ra 0$ holding the boosted lightcone momentum 
density\ ${\lambda^2\varepsilon_{p+1}\over 2}\equiv Q_p$ fixed 
\cite{Singh:2012un}. These describe strongly coupled Yang-Mills 
theories with constant energy flux $T_{++}$.\\
The Einstein metric $ds_E^2 = e^{-\Phi/2} ds_{st}^2$, 
upon dimensional reduction on $S^{8-p}$ and the $x^+$-direction 
(compactified), gives rise to hyperscaling violating spacetimes\ 
$ds^2=r^{2\theta/d} (-{dt^2\over r^{2z}} + {dx_i^2+dr^2\over r^2})$, 
where $t\equiv x^-$, with nontrivial Lifshitz ($z$) and hyperscaling 
violating ($\theta$) exponents 
$\theta={p^2-6p+7\over p-5} ,\ z={2(p-6)\over p-5}$\ (see 
\cite{Singh:2012un}).

\noindent We now calculate entanglement entropy for a strip 
subsystem: 
since this is a non-static spacetime, we use the covariant formulation 
\cite{HRT}, following \cite{Narayan:2012ks} for the conformal case. 
We consider a strip along the energy flux direction, with width 
along some $x_i$-direction, corresponding to a spacelike subsystem. 
The subsystem is specified by\
$0\leq x_1\leq l,\ (x^+,x^-)=(\al y, -\beta y)$, with the range of 
$y$ and the other $x_i$ coordinates being $(-\infty,\infty)$\ (with 
regulated lengths\ $L_i\gg l$).
Then the entanglement entropy (\ref{EEMth}) (\ref{EEMst}) simplifies to
\be\label{SAspacelike}
S_A 
\ \sim\ {V_{p-1} R_p^{7-p}\over G_{10}} \int^{r_0}_{r_*} dr r\ 
\sqrt{2\beta + \al {G_{10}Q\over r^{7-p}}\ }\ 
\sqrt{1+{r^{7-p}\over R_p^{7-p}} (x_1')^2\ }\ ,
\ee
where $r_0$ is the UV cutoff and $r_*$ is the turning point for the 
extremal surface $x_1(r)$. The lightlike limit $\beta=0$ corresponds 
to constant-$x^-$ (null time) slicing: this is natural from the point 
of view of constant time slices in the lower dimensional theory 
obtained by $x^+$-dimensional reduction, with time defined as 
$t\equiv x^-$. However such a null entanglement entropy is tricky to 
define ab initio. We therefore study spacelike subsystems on a 
constant time slice corresponding to $\al=\beta=1$: the scaling 
estimates below for $l(r_*)$ and $S_A^{finite}$ are as in the null case.
Analysing (\ref{SAspacelike}) gives the extremal surface and 
associated entanglement 
\be\label{EEspacelike} 
{dx_1\over dr} = {A R_p^{7-p}\over r^{8-p} \sqrt{2 + {G_{10}Q\over r^{7-p}} 
- {A^2 R_p^{7-p}\over r^{9-p}} }}\ ,\quad 
S_A\ \sim\ {V_{p-1} R_p^{7-p}\over G_{10}} \int^{r_0}_{r_*} dr r\ 
{2 + {G_{10}Q\over r^{7-p}} \over
\sqrt{2 + {G_{10}Q\over r^{7-p}} - {A^2 R_p^{7-p}\over r^{9-p}} }}\ .
\ee
For $Q=0$, these expressions reduce to those for the ground state: 
recall that for the non-conformal $Dp$-branes, the gauge coupling 
$g_{YM}^2=g_s{\al'}^{(p-3)/2}$ is dimensionful and the entanglement 
entropy for strip-subsystems has the form \cite{Ryu:2006ef,Barbon:2008ut}
\be\label{EEgndnonconf}
S_A = N_{eff}(\epsilon) {V_{d-2}\over\epsilon^{d-2}} - 
c_d N_{eff}(l) {V_{d-2}\over l^{d-2}}\ ,\qquad N_{eff}(\epsilon)=N^2 
\Big({g_{YM}^2N\over \epsilon^{p-3}}\Big)^{{p-3\over 5-p}}.
\ee
The two terms here again reflect the local area law and the finite 
cutoff-independent piece, but with a scale-dependent number of degrees 
of freedom $N_{eff}$\ \cite{Barbon:2008ut}. The cutoff 
$\epsilon\equiv {1\over u_0}$ here is written in terms of the 
non-conformal $Dp$-brane supergravity radius/energy variable\
$u={r^{(5-p)/2}\over R_p^{(7-p)/2}}$\ 
introduced in \cite{Peet:1998wn}.\ Thus 
${1\over\epsilon} = {r_0^{(5-p)/2}\over R_p^{(7-p)/2}}$.\\
With $Q\neq 0$, from (\ref{EEspacelike}), the divergent ultraviolet 
behaviour of $S_A$ (large $r$) is seen to be the same as for the 
ground state
\be
S_A^{div}\ \sim\ {V_{p-1} R_p^{7-p}\over G_{10}} r_0^2\ 
=\ N_{eff}(\epsilon) {V_{d-2}\over\epsilon^{d-2}}\ ,
\ee
in accord with the expectation that the excited state leaves 
unaffected the short distance behaviour, while introducing long 
range correlations.
The finite part can be estimated by realising as in \cite{Narayan:2012ks} 
that the turning point can be approximated for large $Q$ 
as~~~${G_{10}Q\over r_*^{7-p}}\  \sim\ {A^2 R_p^{7-p}\over r_*^{9-p}}$~,\
which then, using (\ref{EEspacelike}) alongwith $l\equiv \Delta x_1$, 
gives the scaling estimates
\be
l\sim\ {R_p^{7-p\over 2}\over r_*^{5-p\over 2}}\ ,\qquad\ 
S_A^{finite}\ \sim\ {V_{p-1}\sqrt{Q} \over (3-p) \sqrt{G_{10}}} 
{R_p^{7-p}\over r_*^{(3-p)/2}}\ \sim\ 
{V_{p-1}\sqrt{Q} \over (3-p) \sqrt{G_{10}}} 
{R_p^{(7-p)^2/(2(5-p))}\over l^{(p-3)/(5-p)}}\ .
\ee
Using the expressions above for $R_p,G_{10}$ etc in terms of gauge 
theory parameters $g_{YM},N$, this finite part can be simplified and 
recast as ($p\neq 3$)
\bea
&& S_A^{finite}\ \sim\ {1\over 3-p} {V_{p-1}\sqrt{Q} \over l^{(p-3)/2}} 
N \left({g_{YM}^2N\over l^{p-3}}\right)^{{p-3\over 2(5-p)}}\ = 
{\sqrt{N_{eff}(l)}\over 3-p}\ {V_{p-1}\sqrt{Q} \over l^{(p-3)/2}}\ ,\nonumber\\
&& \qquad\qquad
N_{eff}(l) = N^2 \left({g_{YM}^2N\over l^{p-3}}\right)^{{p-3\over 5-p}} \ .
\eea
$N_{eff}(l)$ is the scale-dependent number of degrees of freedom 
(\ref{EEgndnonconf}) involving the dimensionless coupling at scale $l$.
Recalling that $Q$ is the energy density in $p+1$-dim, the second factor 
is recognized as the natural dimensionless combination of $V_{p-1}, Q, l$, 
given the expectation that the entanglement is proportional to $V_{p-1}, 
\sqrt{Q}$.  Then the finite part suggests that these plane wave states 
are an effective chiral subsector in the gauge theory.
Along the lines of (\ref{QtildeQconf}), a scale-dependent redefinition\ 
$Q= {\tilde Q} N_{eff}(l)$ can be devised to recast the finite part as\ 
$\sim N_{eff}(l) {V_{p-1}\sqrt{{\tilde Q}} \over l^{(p-3)/2}}$ . The 
energy-momentum density ${\tilde Q}$ is then the energy-momentum 
density per nonabelian degree of freedom, but involving the 
dimensionless coupling at scale $l$.

It is interesting to study some specific Dp-brane theories in
particular comparing with their UV/IR conformal phases as in their
phase diagram \cite{Itzhaki:1998dd}. This vindicates the intuition that 
the finite part of entanglement decreases under renormalization group 
flow.

\noindent {\bf D2-M2:}\ \ We have
\be
S_A^{fin}\ \sim\ V_1\sqrt{l} \sqrt{Q}\ \sqrt{{N^2\over (g_{YM}^2Nl)^{1/3}}}\ 
\ (D2)\ ;\qquad
S_A^{fin}\ \sim\ V_1\sqrt{l} \sqrt{Q}\ \sqrt{N^{3/2}}\ \ (M2)\ .
\ee
We see as expected that $S_A^{fin}$ for the plane wave states is 
greater than that for the ground states for the same strip geometry.
Noting the IIA regime of validity\ 
$g_{YM}^2N^{1/5}\ll {r\over\al'} \ll g_{YM}^2N$\ \cite{Itzhaki:1998dd} 
for the turning point $r_*$, we find\ $1\ll g_{YM}^2Nl_{D2}\ll N^{6/5}$.\ 
Thus $N^{3/2}\ll {N^2\over (g_{YM}^2Nl)^{1/3}}\ll N^2$, \ie\ the 
finite entanglement in the D2-supergravity phase is in between the 
free 2+1-dim SYM phase (UV) and the M2-phase (IR), consistent with 
the expectations of the thinning of degrees of freedom under 
renormalization group flow. This also suggests that the free 
2+1-dim SYM entanglement in these plane wave excited states is possibly\
$\sim V_1\sqrt{l} \sqrt{Q}\ \sqrt{N^2}$.

\noindent {\bf D4-M5:}\ \ We have
\be
S_A^{fin}\ \sim\ -{V_3\sqrt{Q}\over \sqrt{l}}\ 
\sqrt{N^2 {g_{YM}^2N\over l}}\ \ (D4)\ ;\qquad
S_A^{fin}\ \sim\ -\sqrt{Q} {V_4\over l}\ \sqrt{N^3}\ \ (M5)\ .
\ee
The finite parts are actually the same expression for both the 
D4-supergravity and M5-phases, if we recognize that the D4-branes 
are M5-branes wrapping the 11th circle (size $R_{11}=g_sl_s=g_{YM}^2$) 
and $V_4=V_3 R_{11} ,\ Q_{D4}=Q_{M5}R_{11}$. The IIA regime of validity\ 
${1\over g_{YM}^2N}\ll {r_*\over\al'}\ll {N^{1/3}\over g_{YM}^2}$\ 
implies\ $1\ll {g_{YM}^2N\over l}\ll N^{2/3}$. The free 5d SYM 
entanglement is possibly\ $-{V_3\sqrt{Q}\over \sqrt{l}} \sqrt{N^2}$.

\noindent {\bf D1:}\ \ We have\ 
$S_A^{fin}\ \sim\ l \sqrt{Q}\ \sqrt{{N^2\over (g_{YM}^2Nl^2)^{1/2}}}$ ,\
which grows as $\sqrt{l}$ (including the $l$-dependence in $N_{eff}$).
The regime of validity\ $g_{YM}N^{1/6}\ll {r_*\over\al'}\ll g_{YM}\sqrt{N}$\ 
implies\ $1\ll g_{YM}^2Nl^2\ll N^{4/3}$.

\subsection{Strip orthogonal to wave}

The subsystem $A$ lying on a constant time-$t$ slice is\ 
$\Delta x^+=-\Delta x^- = {l\over\sqrt{2}}$ with\ $-\infty<x_i<\infty$.
The (covariant) entanglement entropy functional for the extremal 
surface $x^+(r), x^-(r)$ is
\be
S_A\ \sim\ {V_{p-1}R_p^{7-p}\over G_{10}}\int dr r 
\sqrt{1-{2r^{7-p}\over R_p^{7-p}} (\del_rx^+)(\del_rx^-) + 
{G_{10}Q\over R_p^{7-p}} (\del_rx^+)^2\ }\ ,
\ee
giving
\be
S_A\ \sim\ {V_{p-1}R_p^{7-p}\over G_{10}} \int_{r_*}^{r_0} dr r\ {AB \over 
\sqrt{A^2B^2 - 2B {r^{p-9}\over R_p^{7-p}} + Q {r^{2(p-8)}\over R_p^{7-p}} }}\ ,
\ee
\be
{\Delta x^+\over 2} = \int_{r_*}^{r_0} {dr\over \sqrt{A^2B^2r^{16-2p} 
+ {G_{10}Q\over R_p^{7-p}} - {2Br^{7-p}\over R_p^{7-p}}}}\ , \ \ \
{\Delta x^-\over 2} = \int_{r_*}^{r_0} {dr\  ({G_{10}Q\over r^{7-p}}-B)
\over \sqrt{A^2B^2r^{16-2p} 
+ {G_{10}Q\over R_p^{7-p}} - {2Br^{7-p}\over R_p^{7-p}}}}\ .\nonumber
\ee
This structure is similar to the conformal case \cite{Narayan:2012ks}.
Large width $l$ is obtained only when the denominator function\ 
$A^2B^2r_*^{16-2p} + {G_{10}Q\over R_p^{7-p}} - {2Br_*^{7-p}\over R_p^{7-p}}$
acquires a double zero at $r=r_*$:\ this gives\ 
$B={8-p\over 9-p} {G_{10}Q\over r_*^{7-p}}~,\ 
A^2={(7-p)(9-p)\over (8-p)^2} {1\over G_{10}Q R_p^{7-p} r_*^2}$ ,\
for which $\Delta x^+, \Delta x^-$ both become positively divergent, 
incommensurate with a spacelike subsystem. As for the conformal case, 
this suggests a phase transition due to an upper bound on the width: 
no connected surface corresponding to a spacelike subsystem exists 
beyond this critical width, which has a scaling estimate\ 
$l_c\sim R_p^{7-p\over 2}/r_{*,c}^{5-p\over 2} \sim
(g_{YM}^2N)^{{p-3\over 2(7-p)}} ({Q\over N^2})^{p-5\over 2(7-p)}$.
There are multiple length scales here unlike the conformal case, 
constraining the phase transition structure: \eg\ for D2-branes, 
the IIA supergravity regime of validity for $r_{*,c}$ (and so $l_c$) 
gives\ ${(g_{YM}^2N)^3\over N^2} \lesssim Q \lesssim (g_{YM}^2N)^3 N^2$.

\subsection{NS5-brane plane waves}

NS5-branes in certain decoupling limits have been argued to be 
dual to nonlocal 6-dim ``little string'' theories \cite{Berkooz:1997cq} 
(see \eg\ \cite{Maldacena:1997cg,Aharony:1998ub,Kutasov:2001uf,
Giveon:1999px,LST} for further studies). 
Here we identify plane wave states in these theories. 
Starting with the finite temperature NS5-brane system before
decoupling
\be
ds^2=-(1-{r_0^2\over r^2}) dt^2 + \Big(1+{N\al'\over r^2}\Big) 
\left( {dr^2\over 1-{r_0^2\over r^2}} + r^2d\Omega_3^2\right) 
+ \sum_{i=1}^5dy_i^2 ,\quad e^{2\Phi}=g_s^2 \Big(1+{N\al'\over r^2}\Big) ,
\ee
defining lightcone coordinates\ $t={x^++x^-\over\sqrt{2}} ,\
x_5={x^+-x^-\over\sqrt{2}}$~, after a lightlike boost\ 
$x^\pm\ra \lambda^{\pm 1} x^\pm$\ with parameter $\lambda$, we  
identify a new double scaled zero-temperature decoupling 
limit that results in plane wave like excited states in the little 
string theory on the NS5-branes, distinct from the Hagedorn temperature 
limit \cite{Maldacena:1997cg}. The temperature parameter above is\ 
$r_0^2=G_{10}\mu$.\ 
Under the double scaling limit\ $\lambda\ra\infty ,\ g_s\ra 0$, with\ 
${\lambda^2 g_s^2 \mu\over 2} \equiv Q$ fixed (with dimensions of 
boundary energy density), the near horizon spacetime becomes
\be
ds_{st}^2 = -2dx^+dx^- + {Q{\al'}^4\over r^2} (dx^+)^2 
+ \sum_{i=1}^4dy_i^2 + N\al' {dr^2\over r^2} + N\al' d\Omega_3^2\ ,\qquad
e^{2\Phi} = g_s^2 {N\al'\over r^2}\ ,
\ee
an asymptotically linear dilaton background with ``normalizable'' null 
deformation (vanishing as\ $r\ra\infty$).
It can be checked independently that this is a solution to the NS-NS 
sector spacetime equations: the only new contribution here to the 
string frame equations describing the NS5-brane is\ 
$R_{++}=-2\nabla_+\nabla_+\Phi=+2\Gamma_{++}^r\del_r\Phi
=-{2Q{\al'}^3\over Nr^2}$ , which is consistent. This is another way 
to find this solution, and is consistent with S-duality of the 
D5-brane plane waves earlier (for Type IIB). These lightlike 
deformations are likely $\al'$-exact and supersymmetric.

Dimensionally reducing the Einstein metric on $S^3$ and the 
$x^+$-direction (compactified) gives\ 
$ds^2_{E,6d}\sim r^{1/2}(-{r^2\over Q{\al'}^4}dt^2+dy_i^2+N\al'{dr^2\over r^2})$,
\ which is not in hyperscaling violating form.

The double scaling limit here and the resulting NS5-brane plane wave
excited states appear distinct from previous double scaled limits of
little string theory \cite{Giveon:1999px}: it will be interesting 
to explore these NS5-brane or little string excitations further.

Here we briefly discuss holographic entanglement entropy for these 
theories: for a spacelike strip-subsystem of width $l$, as before we have\ 
$S_A\ \sim\ {V_4  N\al' \over G_{10}} \int dr r 
\sqrt{2 + {Q{\al'}^4\over r^2}} \sqrt{1 + {r^2\over N\al'} (x')^2}$~,
giving\
${dx\over dr} = {\sqrt{N\al'}\over r} {A\sqrt{N\al'}\over 
\sqrt{2r^4 + Q{\al'}^4 r^2 - A^2 N\al'}}\ ,\ \ 
S_A \sim {V_4  N\al' \over G_{10}} \int {dr r^3 
(2 + {Q{\al'}^4\over r^2})\over \sqrt{2r^4 + Q{\al'}^4 r^2 - A^2 N\al'}}$ ,
and so (with $U={r\over\al'}$) the scaling estimates\ 
$l\sim \sqrt{N\al'} ,\ \ 
S_A^{UV} \sim {V_4  N\al' \over g_s^2} U_0^2 ,\ \ 
S_A^{finite} \sim {V_4N\sqrt{Q}\over g_s^2} U_*$.\
We see that (as for the ground state \cite{Ryu:2006ef,Barbon:2008ut}) 
$l$ degenerates, being fixed to the scale of nonlocality (independent 
of $r_*$) inherent in these theories.

\newpage
\noindent {\small {\bf Acknowledgments:} I thank S. Trivedi for 
discussions and T. Takayanagi for comments on a draft. This work is 
partially supported by a Ramanujan Fellowship, DST, Govt of India.}

\end{document}